\documentstyle[aps,prb,epsfig]{revtex}
\begin{document}
\twocolumn[\hsize\textwidth\columnwidth\hsize\csname
@twocolumnfalse\endcsname
\title{Justification of the effective fractional Coulomb energy and
the extended Brinkman-Rice picture $^{\ast}$}
\author{Hyun-Tak Kim $^{\ast\ast}$}
\address{Telecom. Basic Research Lab., ETRI, Taejon 305-350, Korea }
\maketitle{}
\begin{abstract}
In order to calculate the effective mass of quasiparticles for a
strongly correlated metallic system in which the number of
carriers, $n$, is less than that of atoms (or lattices), $l$, the
metallic system is averaged by one effective charge per atom over
all atomic sites. The effective charge of carriers,
$<e>=(n/l)e={\rho}e$, and the on-site effective Coulomb repulsion,
$U={\rho}^2{\langle}{\frac{e^2}{r}}{\rangle}\equiv{\rho}^2{\kappa}U_c$,
are defined and justified by means of measurement. ${\rho}$ is
band filling and $\kappa$ is the correlation strength. The
effective mass, $\frac{m^*}{m}=\frac{1}{1-{\kappa}^2{\rho}^4}$,
calculated by the Gutzwiller variational theory, is regarded as
the average of the true effective mass in the Brinkman-Rice (BR)
picture and is the effect of measurement. The true effective mass
has the same value irrespective of $\rho$, and is not measured
experimentally. The true correlation strength of the BR picture is
evaluated as $0.90<\kappa_{BR}<1$ for Sr$_{1-x}$La$_x$TiO$_3$ and
$0.92<\kappa_{BR}<1$ for YBCO. High-$T_c$ superconductivity for
YBCO is attributed to the true effective mass caused by the large
$\kappa_{BR}$ value. The effective mass explains the
metal-insulator transition of the first order on band filling.
Furthermore, the pseudogap is predicted in this system.
\\ \\
\end{abstract}
]
\newpage
\section{INTRODUCTION}
Ever since a correlation-driven metal-Mott-insulator
transition$^{1}$ of the first order (called the "Mott transition")
was predicted in 1949, the Mott transition has never been
completely solved. The Hubbard model$^{2}$, one of the theories
explaining the Mott transition, showed the second-order
metal-insulator transition(MIT) by increasing the on-site Coulomb
repulsion $U$, whereas the Brinkman-Rice (BR) picture$^{3}$, based
on Gutzwiller's variational theory$^{4}$, explained the
first-order transition at $U/U_c$=1 for the first time. The
infinite dimensional($d=\infty$) Hubbard model suggested that the
MIT occurs while going from $U/W$ to $U_c/W$, and that the
metallic side exhibits the BR picture.$^{5}$ Here $W$ is the band
width. The $d=\infty$ Hubbard model and the BR picture were built
on a metallic system with an electronic structure of one electron
per atom. However the metallic system which Gutzwiller assumed is
a general one without limiting the number of electrons per atom .
Furthermore, observations and current understanding of the MIT
driven by correlation effects were reviewed.$^{6,7}$ Validity of
the BR picture and the Mott transition was also discussed by
theorists, which is because the BR picture has not been proved
experimentally.$^{8,9}$

Experimentally, for the coefficient $\gamma$ of the heat capacity
of Sr$_{1-x}$La$_x$TiO$_3$ (SLTO) which is proportional to the
effective mass, the sharp first-order transition was observed
between $x$=0.95 and 1.$^{10,11}$ This transition was regarded as
the Mott transition on band filling, which could not be explained
by the above mentioned Hubbard models and the BR picture. This is
because the structure of the metallic system of SLTO is not one
electron per atom. Hence, when the number of carriers is less than
the number of atoms, the MIT of first-order on band filling has
remained a theoretical problem. In particular, the metallic
property near the Mott transition is still
controversial.$^{12-14}$ Furthermore the effective mass of
quasiparticles as a function of band filling is necessary to take
account of high $T_c$ superconductivity.$^{15}$

Recently, an insight on both the effective mass and the Mott
transition was proposed on the basis of
$U={\rho}^2{\langle}{\frac{e^2}{r}}{\rangle}$ and experimental
data.$^{16}$ However the justification of
$U={\rho}^2{\langle}{\frac{e^2}{r}}{\rangle}$ and exact
calculations for the effective mass were not given. Thus the
problem of the Mott transition on band filling remains unanswered.

In this paper, when the number of carriers is less than the number
of atoms, the necessity of the development of $<e>={\rho}e$ is
given by introducing the instability of the charge-density-wave
potential energy, and the justification of $<e>={\rho}e$ and
$U={\rho}^2{\langle}{\frac{e^2}{r}}{\rangle}$ is given by means of
measurement. The effective mass of quasiparticles is exactly
calculated on the basis of the Gutzwiller variational theory.
Furthermore, true correlation strengths of the BR picture, not
given in reference 16, are evaluated for Sr$_{1-x}$La$_x$TiO$_3$
and YBCO.

\section{EFFECTIVE FRACTIONAL COULOMB ENERGY}
A strongly correlated metallic system of the $s$ band structure is
assumed for a $d=\infty$ dimensional simple-cubic lattice. Let $n$
and $l$ be the number of electrons (or carriers) and the number of
atoms (or lattices), respectively. In the case of one electron per
atom, $i.e.$, $n=l$, the metallic system is a metal and the
existence probability ($P=n/l={\rho}$= band filling) of electrons
on nearest-neighbor sites is one. The on-site Coulomb repulsion is
always given by $U=U'{\equiv}{\langle}{\frac{e^2}{r}}{\rangle}$
and an electron on an atom has a spin, as shown in Fig. 1 (a).
However, in the case of $n<l$ by doping an element to a base
insulator or metal, $U$ is determined by probability. The
metallic system is quite complicated, as shown in Fig. 1 (b).
Four types of regions for the system can be distinguished as
possible extreme examples. Region A in Fig. 1 (b) has no
electrons on its atomic sites, which corresponds to a normal
insulator. Region C has the metallic structure of one electron
per atom. Regions B and D have a charge-density-wave (CDW)
structure unlike the assumed cubic. This must be regarded as a
CDW insulator with the CDW-energy gap$^{17}$ depending on the
local CDW-potential energy.$^{17,18}$ Moreover, even when one
electron on an atom is removed in Fig. 1 (a), both
nearest-neighbor sites of the atom without an electron and the
atom itself also are regarded as region B. The CDW-potential
energy is defined by $V_{CDW}=-E_p(Q_i-Q_j)^2$, where $Q_i$ and
$Q_j$ are charges irrespective of spins on $i$ and $j$ sites,
respectively. $E_p$ is the small-polaron binding energy which is
a constant of electron-phonon coupling. The potential energy is
derived by breathing-mode distortion (or frozen-breathing mode)
due to a charge disproportionation (${\delta}Q=Q_i-Q_j$) between
nearest-neighbor sites. The CDW-energy gap is regarded as a
pseudogap (or pinned CDW gap) for this system. Here ${\delta}Q$
is called the electronic structure factor in real-space.
${\delta}Q$=0 indicates that the electronic structure is
metallic, while ${\delta}Q\ne$0 suggests that it is insulating.
In addition, $V_{CDW}$ can also be expressed in terms of band
filling as follows: $V_{CDW}=-E_p(1-\rho)^2=-E_p{\mu}^2$, where
$\mu=1-\rho=n_b/l$~(or $n_b=l-n$), $n_b$ is the number of bound
charges bounded by $V_{CDW}$.$^{18}$ As $\rho$ increases, $\mu$
decreases. This indicates that the pseudogap in the CDW
insulating phase decreases as the metallic phase increases,
because the pseudogap depends upon $V_{CDW}$.

Conversely, the pseudogap, as observed by optical
methods,$^{19-22}$ photoemission spectroscopy,$^{23}$ and heat
capacity measurement$^{24,25}$ for high-$T_c$ superconductors,
necessarily occurs if the number of carriers is less than the
number of atoms. Therefore, the metallic system is inhomogeneous
and cannot be self-consistently represented in $k$-space, $i.e.$,
$U$ and the density of states of the system are not given. Thus,
in order to overcome this difficulty, charges on atoms in the
metal phase (region C) have to be averaged over sites, as shown
in Fig. 1 (c). The on-site effective charge is
$Q_i=Q_j=<e>=(n/l)e={\rho}e$, on average over sites. The metallic
system, then, is a metal, because $V_{CDW}=0$ due to
${\delta}Q$=0. This system is analogous to the case of one
electron per atom, such as Fig. 1 (a). Therefore, the effective
Coulomb energy is given by $U={\rho}^2U^{\prime}$, as shown in
Fig. 1 (c).

To illustrate the physical meaning of $P=(<e>/e)=\rho<1$, then
$\mu<<1$, $\rho=1-\mu<1$. For the CDW insulating side, $\mu<<1$
indicates that there are a few doubly occupied atoms such as in
region D, while, for the metallic side, $\rho<1$ indicates that a
number of carriers exist. In other words, this suggests that
region C is extremely wide.

The insulating and metallic sides correspond to two phases, which
are attributed to the metal-insulator instability(or instability
of the CDW-potential energy)$^{18}$ at $\rho$=1 (half filling).
This indicates that metals with the electronic structure of Fig.
1 (a) are not synthesized. Junod et al.$^{26}$ suggested that even
the best samples are not 100$\%$ superconducting, which supports
the metal-insulator instability$^{18}$. Thus synthetic metals
composed of several atoms always have the electronic structure of
two phases such as Fig. 1 (b), which is a necessity of the
development of $<e>={\rho}e$.

To justify the fractional charge of carriers of the metallic
system with two phases, the concept of measurement is necessary.
In the case of measuring charge of carriers in the metallic system
of Fig. 1 (b), the measured charge becomes the effective
fractional one, $<e>={\rho}e$, with the average meaning of the
system because experimental data for the metallic system are
expectation values of statistical averages. In the case of not
measuring charge, the charge of carriers in region C in Fig. 1 (b)
remains the true elementary one not observed by means of
measurement.

Accordingly, for the metallic system of the $n{\le}l$ case, the
effective fractional Coulomb energy, $U={\rho}^2U^{\prime}$, is
defined by using $<e>={\rho}e$ and justified by means of
measurement.

Furthermore, $U={\rho}^2U^{\prime}$, deduced on the basis of the
CDW concept, can be applied to even cuprate systems such as
La$_{1-x}$Sr$_x$Cu$_2$O$_4$(LSCO). This is because the MIT with
$x$ from the antiferromagnetic insulator, La$_2$CuO$_4$, to a
metal of LSCO is continuous although there is a controversy as
for the MIT.

\section{CALCULATION OF THE EFFECTIVE MASS}
For the metallic system regarded as a real synthetic metal with an
electronic structure such as Fig. 1 (b), the effective mass of
quasiparticles needs to be calculated. Hamiltonians of the
metallic system can be considered as follows. Hamiltonian, $H$, is
given by

\begin{eqnarray}
H &=& H_1 + H_2,\\ H_1 &=&
\sum_{k}(a_{k\uparrow}^{\dagger}a_{k\uparrow}+
a_{k\downarrow}^{\dagger}a_{k\downarrow}){\epsilon}_k+ U
\sum_{g}a_{g\uparrow}^{\dagger}a_{g\downarrow}^{\dagger}a_{g\downarrow}a_{g\uparrow},
\\
H_2 &=& - \sum_{i,j}E_p(Q_i-Q_j)^2,
\end{eqnarray}
where $a_{k\uparrow}^{\dagger}$ and $a_{g\uparrow}^{\dagger}$ are
the creation operators for electrons in the Bloch state ${k}$ and
the Wannier state ${g}$, respectively, and ${\epsilon}_k$ is the
kinetic energy when $U$=0. $H_1$ and $H_2$ are Hamiltonians of the
metallic region C and the CDW-insulator regions B and D,
respectively. In the case of Fig. 1 (a) and (c), the Hamiltonian
is reduced to $H_1$ because $H_2$ disappears due to ${\delta}Q=0$,
and the on-site Coulomb energy is given by $U={\rho}^2U^{\prime}$.
$H_1$ is consistent with the Hamiltonian used in the Gutzwiller
variational theory$^{4}$.

In order to calculate the effective mass of quasiparticles and the
ground-state energy for a strongly correlated metallic system, the
Gutzwiller variational theory$^{4,27-29}$ is used. $H_1$ is
supposed to describe the metallic system. The wave function is
written as
\begin{eqnarray}
\vert\Psi\rangle={\eta}^{\bar\nu}{\vert\Psi}_0\rangle,
\end{eqnarray}
where ${\vert\Psi}_0\rangle$ is the wave function when $U=0$,
${\bar\nu}$ is the number of doubly occupied atoms, and
$0<{\eta}<1$ is variation.$^{4}$ The expectation value of $H_1$
is regarded to be
\begin{eqnarray}
{\langle}H{\rangle}=\frac{{\langle\Psi\vert\sum_{ij}\sum_{\sigma}t_{ij}a_{i\sigma}^{\dagger}a_{j\sigma}
 \vert\Psi\rangle}+{{\langle\Psi\vert}U{\sum_i\rho_{i\uparrow}\rho_{i\downarrow}\vert\Psi\rangle}}}
 {\langle\Psi\vert\Psi\rangle}.
\end{eqnarray}
The second part of the equation is simply given by
${{\langle\Psi\vert}U{\sum_i\rho_{i\uparrow}\rho_{i\downarrow}\vert\Psi\rangle}}$=$U\bar\nu$
because ${\vert\Psi}_0\rangle$ is an eigenstate of the number
operator ${\sum_i\rho_{i\uparrow}\rho_{i\downarrow}}$. The first
part is dealt with by assuming that the motion of the up-spin
electrons is essentially independent of the behavior of the
down-spin particles (and $vice ~versa$). By minimization with
respect to ${\eta}$, Gutzwiller obtained an extremely simple
result for the ground-state energy, namely,
\begin{eqnarray}
E_g/l=q_{\uparrow}({\bar\nu},\rho_{i\uparrow},\rho_{i\downarrow}){\bar\epsilon_{\uparrow}}+q_{\downarrow}
({\bar\nu},\rho_{i\uparrow},\rho_{i\downarrow}){\bar\epsilon_{\downarrow}}+U{\bar\nu}.
\end{eqnarray}
Here,
\begin{eqnarray}
{\bar\epsilon_{\sigma}}=l^{-1}{\langle\Psi\vert\sum_{ij}t_{ij}a_{i\sigma}^{\dagger}a_{j\sigma}
 \vert\Psi\rangle}=\Sigma_{k<k_F}\epsilon_k<0
\end{eqnarray}
is the average energy of the $\sigma$ spins without correlation
and $\epsilon_k$ is the kinetic energy in $H_1$, with the zero of
energy chosen so that $\Sigma_k\epsilon_k$=0.
${\bar\epsilon_{\uparrow}}$ is equal to
${\bar\epsilon_{\downarrow}}$.

The discontinuities, $q_{\sigma}$, in the single-particle
occupation number at the Fermi surface are given by
\begin{eqnarray}
q_{\sigma}=\frac{\left(\sqrt{(\rho_{\sigma}-{\bar{\nu}})(1-\rho_{\sigma}-\rho_{-\sigma}+
{\bar\nu})}+\sqrt{(\rho_{-\sigma}-{\bar\nu}){\bar\nu}}\right)^2}{\rho_{\sigma}(1-\rho_{\sigma})},
\end{eqnarray}
where $\rho_{\sigma}=\frac{1}{2}\rho$, $0<{\rho\le}1$ and
$\rho_{\uparrow}=\rho_{\downarrow}$.$^{27,28}$ This, calculated
by Ogawa et. al.$^{27}$ who simplified the Gutzwiller variational
theory, is in the context of the Gutzwiller variational theory.
Eq. (8) is a function of $\rho_{\sigma}$ and $\bar{\nu}$
irrespective of the quantity of charges. This can be analyzed in
two cases of $\rho$=1 and $0<\rho<$1, because Gutzwiller did not
limit the number of electrons on an atom for the metallic system.

In the case of $\rho$=1,
\begin{eqnarray}
q_{\sigma}=8\bar\nu(1-2\bar\nu).
\end{eqnarray}
This was described in the BR picture.

In the case of $0<\rho<1$, two kinds of $q_{\sigma}$ can be
considered. One is Eq. (8) when the electronic structure is
${\delta}Q\ne$0, as shown in Fig. 1 (b). However, Eq. (8) can not
be applied to the metallic system, as mentioned in the above
section. The other is Eq. (9) when the electronic structure is
${\delta}Q$=0, as shown in Fig. 1 (c). Eq. (9) is obtained from
substituting ${\rho}_{\sigma}$ in Eq. (8) with
${\rho}^{\prime}_{\sigma}$. Here,
${\rho}^{\prime}=(n^{\prime}/l)=1$ and
${\rho}^{\prime}_{\sigma}=\frac{1}{2}$, because the number of the
effective charges, $n^{\prime}$, is equal to $l$. It should be
noted that the metallic system with less than one electron per
atom, as shown in Fig. 1 (c), is mathematically consistent with
that with one electron per atom, as shown in Fig. 1 (a).

Although the following calculations were performed by Brinkman and
Rice,$^{3}$ the calculations are applied to the effective mass.
In the case of Fig. 1 (c), by applying Eq. (9) to Eq. (6) and by
minimizing it with respect to $\bar\nu$, the number of the doubly
occupied atoms is obtained as
\begin{eqnarray}
\bar\nu=\frac{1}{4}(1+\frac{U}{8\bar\epsilon})&=&\frac{1}{4}(1-\frac{U}{U_c}),
\nonumber
\\ &=&\frac{1}{4}(1-{\kappa}{\rho}^2),
\end{eqnarray}
where $U_c$=8${\vert\bar\epsilon\vert}$ because of
$\bar\epsilon={\bar\epsilon_{\uparrow}}+{\bar\epsilon_{\downarrow}}<$0,
$U={\rho}^2U^{\prime}$ and $U^{\prime}={\kappa}U_c$.
$0<\kappa{\le}1$ is the correlation strength. By applying Eq. (10)
to Eq. (9) again, the effective mass is given by
\begin{eqnarray}
q_{\sigma}^{-1}=\frac{m^*}{m}&=&\frac{1}{1-(\frac{U}{U_c})^2},
\nonumber
\\ &=&\frac{1}{1-{\kappa}^2{\rho}^4}.
\end{eqnarray}

Although the separate conditions are $0<{\rho}{\le}1$  and
$0<{\kappa}{\le}1$, $m^{\ast}$ is defined under the combined
condition $0<{\kappa}{\rho}^2<1$ and is an average of the true
effective mass in the BR picture for metal phase (region C in Fig.
1 (b)). The effective mass increases as it approaches ${\kappa}$=1
and ${\rho}$=1. For ${\kappa}{\ne}$0 and ${\rho}{\rightarrow}$0,
the effective mass decreases  and, finally, the metallic system
undergoes a normal (or band-type) metal-insulator transition; this
transition is continus. The system at ${\kappa}{\rho}^2=1$ can be
regarded as the insulating state which is the paramagnetic
insulator because ${\bar{\nu}}=0$. At a $\kappa$ value (not one),
the MIT from a metal at a $\rho$ value of just below $\rho$=1 to
the insulator at both $\rho$=1 and $\kappa$=1 is the first-order
transition on band filling. This has been called the Mott
transition by a lot of scientists including author. However, this
is theoretically not the Mott transition which is an first-order
transition from a value of $U$ to $U_c$ in a metal with the
electronic structure of one electron per atom at $\rho$=1, as
given in the BR picture. The Mott transition does not occur in
real crystals because a perfect single-phase metal with $\rho$=1
is not made. Conversely, by hole doping (or electron doping to a
metallic system with hole carriers) of a very low concentration,
the first-order transition from the insulator with $\bar{\nu}$=0
to a metal can be interpreted as an abrupt breakdown of the
balanced critical Coulomb interaction, $U_c$, between electrons.
Then, the $U_c$ value in the insulator reduces to a $U$ value in a
metal phase and an insulating phase produces due to doping of
opposite charges, as shown in Fig. 1(d). This first-order
transition with band filling is very important result found in
this picture, which differs from the continuous (Mott-Hubbard or
second-order) transition by a large $U$ given by the Hubbard
theory.

In order to obtain the expectation value of the energy in the
(paramagnetic) ground state, Eqs. (10) and (11) are applied to Eq.
(6). $E_g$ is given by
\begin{eqnarray}
E_g/l={\bar{\epsilon}}(1-{\kappa}{\rho}^2)^2.
\end{eqnarray}
As $U/U_c={\kappa}{\rho}^2$ approaches one, $E_g$ goes to zero.

In addition, the spin susceptibility in the BR picture is
 replaced by
\begin{eqnarray}
\chi_s&=&{\frac{m^{\ast}}{m}}{\frac{\mu_B^2N(0)}{(1-{\frac{1}{2}}N(0){\kappa\rho}^2U_c\frac{1+\frac{1}{2}
{\kappa\rho^2}}{(1+{\kappa\rho}^2)^2})}},
\end{eqnarray}
where $N(0)$ is the density of states at the Fermi surface and
$\mu_B$ is the Bohr magneton. The susceptibility is proportional
to the effective mass which allows the enhancement of $\chi_s$.

In the above picture, $m^{\ast}$, $E_g/l$ and $\chi_s$ are the
averages of true effective values in region C in Fig. 1 (b) which
is described by the BR picture, and are justified only by means of
measurement. The true effective ones in the BR picture are not
measured experimentally, as mentioned in an above section. In
particular, the magnitude of the true effective mass in the BR
picture has the same value regardless of the extent of region C,
while the measured effective mass depends upon the extent of
region C.

\section{CONCLUSION}
In order to evaluate true correlation strengths of
Sr$_{1-x}$La$_x$TiO$_3$ and YBCO, explanations not given in the
previous paper$^{16}$ are appended. The experimental data of the
heat capacity of Sr$_{1-x}$La$_x$TiO$_3$ was well fitted by Eq.
(11) with $\kappa$=1 for all $x$ below $x=\rho$=0.95. $\kappa$=1
indicates that the effective mass has a constant value although
$\rho$ changes. The increase of $x$ to $x$=0.95 corresponds to the
increase of region C. Because region C is described by the BR
picture, the true correlation strength, $\kappa_{BR}=U/U_c$, of
the BR picture can be found. When it is assumed that the extent of
the metal phase (region C) at $x$=0.95 is closely the same as that
at $\rho$=1, a value of $m^{\ast}/m=1/(1-{\rho}^4)$ at
$x={\rho}$=0.95 and $\kappa$=1 is approximately equal to that of
$m^{\ast}/m=1/(1-{\kappa}_{BR}^2)$ in the BR picture. Then
${\kappa}_{BR}$ = (0.95)$^2$=0.90 is obtained, which indicates
that La$_{1-x}$Sr$_x$TiO$_3$ is very strongly correlated.
Moreover, the decrease of the effective mass from $x$=0.95 to
$x$=0 is the effect of measurement not the true effect.

In the case of YBCO$_{6+\rho}$ at ${\rho}$=0.96, as shown in Fig.
1 (c) and (d) of reference 16, the true correlation strength is
determined as $\kappa_{BR}$=0.92 by the same above method.
High-$T_c$ superconductivity for YBCO is attributed to the true
effective mass caused by the large $\kappa_{BR}$ value. The heat
capacity data measured by Loram et al.$^{24}$ seems to be
explained by the above picture, too.

As another experimental result, the first-order transition at
$x$=0.02 for $h$-BaNb$_x$Ti$_{1-x}$O$_3$ was observed, which may
also be in the context of the above picture.$^{30}$ The validity
of the BR picture is indirectly proved through the above picture.
Furthermore, the combination of the above picture and the BR
picture$^{3}$ is called the extended BR picture defined in
$0<{\kappa}{\rho}^2<1$.

\begin{center}
\noindent{\bf ACKNOWLEDGEMENTS}
\end{center}
I would like to acknowledge Dr. Kwang-Yong Kang for providing the
research environment necessary for this research.

\newpage
\begin{figure}
\vspace{0.1cm} \centerline{\epsfxsize=9.0cm\epsfbox{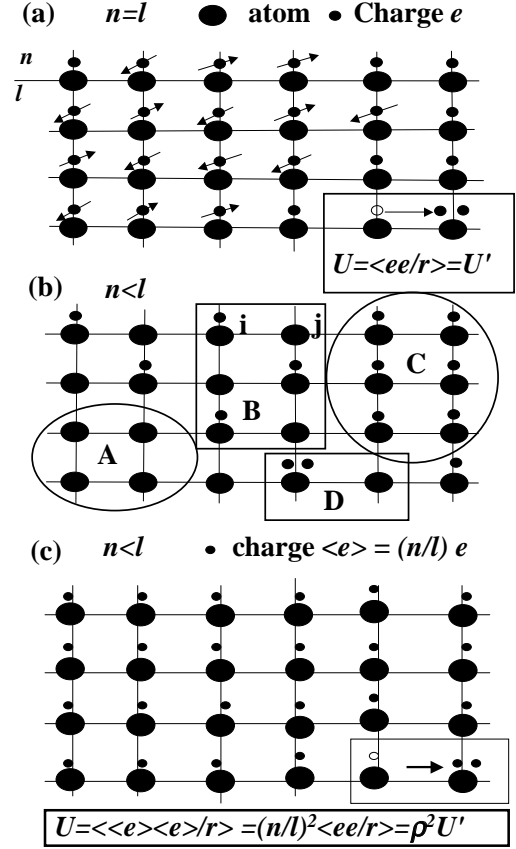}}
\vspace{0.1cm} \caption{(a) In the case of one electron per atom,
the on-site Coulomb repulsion is given by
$U={\langle}\frac{e^2}{r}{\rangle}=U'$. (b) In the $n<l$ case,
the four possible electronic structures are region A (insulator),
region C (metal), and regions B and D (CDW insulator). Here, $n$
corresponds to the number of carriers in region C. (c) The
metallic structure in the case of less than one electron per atom
is shown.
$U=(n/l)^2{\langle}\frac{e^2}{r}{\rangle}={\rho}^2U^{\prime}$.}

\end{figure}


\begin{references}
\bibitem[\ast]{}This was merged in a full-paper presented at NATO Advanced Research
Workshop on New Trends in Superconductivity held at Yalta in
Ukraine at Sept. 16-20 of 2001. Please see cond-mat/0110112.
\bibitem[\ast\ast]{}kimht45@hotmail.com.
\bibitem[\ast\ast]{}htkim@etri.re.kr..

\bibitem{1} N. F. Mott,  Metal-Insulator Transitions, Chapter  3, (Taylor
{\&}  Frances, 2nd edition, 1990).
\bibitem{2} J. Hubbard, (III) ibid {\bf 281}, (1965) 401.
\bibitem{3} W. F. Brinkman and T. M. Rice, Phys. Rev. B{\bf 2}, (1970) 4302.
\bibitem{4} M. C. Gutzwiller, Phys. Rev. 137, (1965) A1726.
\bibitem{5} X. Y. Zhang, M. J. Rozenberg,  and G. Kotliar, Phys. Rev. Lett.
{\bf  70}, (1993) 1666.
\bibitem{6} A. Georges, G. Kotliar, W. Krauth, and M. J.
Rozenberg, Rev. Mod. Phys. 68, (1996) 13.
\bibitem{7} M. Imada, A. Fujimori, and Y. Tokura, Rev. Mod.
Phys. 70, (1998) 1039.
\bibitem{8} T. Moriya, BUTSURI (edited by Physical Society of
Japan), Vol 54, No. 1, (1999) 48 (Japanese).
\bibitem{9} H. Fukuyama, M. Imada, and T. Moriya, BUTSURI
(edited by Physical Society of Japan), Vol 54, No. 2, (1999)
123(Japanese).
\bibitem{10}  Y. Tokura, Y. Taguchi, Y. Okada, Y. Fujishima, T. Arima, K. Kumagai, and Y.
Iye, Phys. Rev. Lett. {\bf 70}, (1993) 2126.
\bibitem{11}   K. Kumagai, T. Suzuki, Y. Taguchi, Y. Okada, Y.  Fujishima, and Y. Tokura,
Phys. Rev. B {\bf 48}, (1993) 7636.
\bibitem{12}  K. Morikawa, T. Mizokawa,  K. Kobayashi, A. Fujimori, H.  Eisaki, S. Uchida,
F. Iga, and Y. Nishihara, Phys. Rev. B {\bf 52}, (1995) 13711.
\bibitem{13} I. H. Inoue, O. Goto, H. Makino, N. E. Hussey, and M. Ishikawa, Phys. Rev.
B {\bf 58}, (1998) 4372.
\bibitem{14}  H. Makino, I.  H. Inoue, M.  J. Rozenberg, I.  Hase, Y. Aiura,  and S. Onari,
Phys. Rev. B {\bf 58}, (1998) 4384.
\bibitem{15} C. C. Tsuei, C. C. Chi, D. M.  Newns, P. C. Pattnaik, and M. Daumling, Phys.
Rev. Lett. {\bf 69}, (1992) 2134.
\bibitem{16} Hyun-Tak Kim, Physica C{\bf 341-348}, (2000) 259:cond-mat/0001008.
\bibitem{17} T. M. Rice and L. Sneddon, Phys. Rev. Lett. {\bf 47}, (1981) 689.
\bibitem{18} Hyun-Tak Kim, Phys. Rev. B {\bf 54}, (1996) 90.
\bibitem{19} T. Timusk and B. Statt, ${\it ~Rep. ~Prog. ~Phys.}$ {\bf 62},
(1999) 61.
\bibitem{20} S. H. Blanton, R. T. Collins, K. H. Kelleher, L. D.
Rotter, Z. Schlesinger, D. G. Hinks, and Y. Zheng, Phys. Rev. B
{\bf 47}, (1993) 996.
\bibitem{21)} M. A. Karlow, S. L. Cooper, A.
L. Kotz, M. V. Kelvin, P. D. Han, and D. A. Payne, Phys. Rev. B
{\bf 48}, (1993) 6499.
\bibitem{22} Hyun-Tak Kim, H. Uwe, and H.
Minami, Advances in Superconductivity VI (Springer-Verlag, Tokyo,
1994), P. 191.
\bibitem{23} H. Ding, T. Yokaya, J. C. Campuzano, T. Takahasi, M.
Randeria, M. R. Norman, T. Mochiku, K. Kadowaki, and J.
Giapinzakis, ${\it Nature}$ {\bf 382}, (1996) 51.
\bibitem{24} J. W. Loram, K. A. Mirza, J. R. Cooper, W. Y. Liang,
and J. M. Wade, Journal of Superconductivity 7, (1994) 243.
\bibitem{25} J. W. Loram, J. L. Luo, J. R. Cooper, W. Y. Liang,
and J. L. Tallon, Physica C{\bf 341-348}, (2000) 831.
\bibitem{26} A. Junod, D. Eckert, T. Graf, G. Triscone, and J.
Muller, Physica C 162-164, (1989) 482.
\bibitem{27} Tohr Ogawa, Kunihiko Kanda, and Takeo Matsubara,
Prog. Theor. Phys. {\bf 53}, (1975) 614.
\bibitem{28} Dieter Vollhardt, Rev. Mod. Phys. {\bf 56}, (1984) 99.
\bibitem{29} Patrick Fazekas, Lecture Notes on Electron
Correlation and Magnetism, Chapter 9, (World Scientific Co.,
1999).
\bibitem{30} M. N. Khan, Hyun-Tak Kim, H. Minami, and H. Uwe,
Materials Letters 47, (2001) 95.
\end{references}
\end{document}